# Nonlinear mode conversion in monodomain magnetic squares

Mikhail Kostylev[1a)], Vladislav E. Demidov[2b)], Ulf-Hendrik Hansen[2], and Sergej O. Demokritov[2]

[1]*School of Physics, University of Western Australia, M013, Stirling Hwy, 6009 Crawley, WA, Australia*

[2]*Institute for Applied Physics, University of Muenster, Corrensstr. 2-4, 48149 Muenster, Germany*

Modifications of spatial distributions of dynamic magnetization corresponding to spin-wave eigenmodes of magnetic squares subjected to a strong microwave excitation field have been studied experimentally and theoretically. We show that an increase of the excitation power leads to a nonlinear generation of long-wavelength spatial harmonics caused by the nonlinear cross coupling between the eigenmodes. The analysis of the experimental data shows that this process is mainly governed by the action of the nonlinear spin-wave damping. This conclusion is further supported by the numerical calculations based on the complex Ginzburg-Landau equation phenomenologicaly taking into account the nonlinear damping.

# I. INTRODUCTION

In the last decades a lot of efforts have been put into experimental and theoretical studies of high-frequency magnetization dynamics in small laterally confined magnetic film structures (see, e.g., [1-20]). This interest is explained by the potential of micrometer- and sub-micrometer-size magnetic structures for applications in integrated data-storage and communication technologies. Besides, a lot of interesting phenomena have been found in such systems, which are of general interest for physics of magnetism. Among them one can mention spin-wave quantization [1-20], spin-wave wells [4,15,17,19], and collective excitations in arrays of magnetic elements [3,20], which were widely addressed in the recent years. Up to very recently these investigations were mainly conducted for the case of relatively small excitation amplitudes providing the linear response of the spin system of ferromagnetic films. This linear dynamics seems to be understood now very deeply for different shapes of magnetic elements and their magnetization conditions including quasi-uniform [1-5,7-9,15,17,19] and non-uniform [6,10-14,16,18] magnetization states.

The progress in the understanding of linear magnetization dynamics on the sub-micrometer scale necessarily leads to the further addressing of more complicated magnetic phenomena in confined film structures appearing at high amplitudes of dynamic magnetization. First, in the magnetic nano-devices, e.g., spin-torque-tranfer nano-oscillators, the amplitudes of dynamic magnetization are essentially large (see, e.g. [21]), which makes the deep understanding of the nonlinear dynamics necessary for their technical applications. Second, investigations on nonlinear dynamics in magnetic systems bring a lot of new information to the physics of nonlinear phenomena in general. Despite the above reasons the number of experimental works in the area of dynamic nonlinear phenomena in microscopic magnetic structures is still rather small due to the evident complexity of such experimental studies. To our knowledge the only dynamic nonlinear phenomenon experimentally observed in small magnetic elements is the nonlinear ferromagnetic resonance [22]. Recently we have suggested the use of low-loss dielectric films of yttrium iron garnet (YIG) in combination with space-resolved Brillouin light scattering (BLS) spectroscopy as a method for experimental modeling of nonlinear magnetization dynamics in small magnetic elements [23]. Using this technique we have observed a modification of the spatial profiles of spin-wave eigenmodes in magnetic squares at large angles of magnetization precession. Here we present a deep analysis of the observed phenomenon based on the complex Ginzburg-Landau equation. We show that the nonlinearity in the system causes a coupling between the eigenmodes leading to the nonlinear energy channeling from the excited mode into other modes. We find that the nonlinear coupling results in nonlinear generation of additional long-wavelength spatial harmonics, which is not typical for systems with attractive nonlinearity. We associate these observations with the influence of the nonlinear magnetic damping.

## II. EXPIREMENT

The samples used in the experiments were patterned using optical lithography and chemical etching from epitaxial single-crystalline YIG films with the thickness of 5.1 µm. The studied elements are squares with the side $w$ = 2 mm. The excitation of dynamic magnetization in the elements was performed by means of a wire antenna with the diameter of 50 µm attached to the square along its middle line. The power of the exciting microwave signal $P$ was varied from 1 mW, providing the linear response of the system, to 200 mW. A static magnetic field of $H$=800-2000 Oe, creating a monodomain magnetization state of the square, was applied in the film plane perpendicularly to the antenna. The frequencies of the spin-wave eigenmodes were determined observing the signal reflected from the antenna by means of a microwave network analyzer. In this way, we obtained the frequencies, at which the absorption of the supplied microwave signal demonstrates maxima (see Fig. 1). The identification of the observed peaks was performed based on the mapping of the profiles of the dynamic magnetization at the frequencies of maximum absorption. The two-dimensional mapping of the dynamic magnetization was done by means of the Brillouin light scattering (BLS) spectroscopy in the forward scattering geometry [24]. For this geometry the BLS intensity is proportional to the square of the precession angle of the magnetisation, $|\varphi|^2$ averaged over the probing laser spot, which in these experiments had a diameter of about $d$=60 μm. A detailed description of the used experimental setup can be found elsewhere [23]. Note, that the absolute value of the opening angle was not experimentally determined due to complexity of the calibration process of the BLS setup. However, the technique gives a possibility to map a distribution of the opening angle over the sample in spin-wave modes with the wave vectors up to $\pi/d$.

It is well known (see, e.g., [15,17]), that the linear eigenmodes of a rectangular magnetic film element represent a combination of two standing waves in the two orthogonal lateral directions. These eigenmodes are usually labeled with two integer indexes, characterizing the number of antinodes in the standing waves in the two directions – quantization indexes. Following this convention we label the peaks in Fig. 1 with a pair of indexes ($n,m$) corresponding to the directions parallel and perpendicular to the static magnetic field, respectively. Due to the specific symmetry of the excitation geometry only eigenmodes with an even number of antinodes in the direction perpendicular to the antenna and with an odd number of antinodes in the direction along the antenna can be excited in our experiments. Correspondingly the peaks in Fig. 1 have even first index $n$ and odd second index $m$.

As already mentioned, the structure of linear eigenmodes of a monodomain magnetic rectangle is well understood. This study is aimed at the investigation of the modification of the spatial structure and the frequencies of the modes occurring when the excitation power is increased. Figure 2 presents an example of such modifications for two modes (2,1) and (4,1). The figure consists of four panels presenting the maps of the square of the precession angle of the magnetization and their two-dimensional Fourier spectra reflecting the distributions of the same value in the wave vector space. For both modes the data are shown for the linear case ($P$=1 mW) and for the strongly nonlinear case ($P$=200 mW). As seen from the figure, in the linear case, the maps demonstrate well-pronounced standing waves with two and four antinodes in the direction parallel to the static field $H$ for the modes (2,1) and (4,1), respectively. Accordingly, the Fourier spectra exhibit peaks at the wave vector components $k_y=\pi/w$ and $k_z=2\pi/w$ and $4\pi/w$, where the $z$- and $y$-axis are aligned parallel and perpendicularly to the applied static magnetic field $H$.

As the excitation power is increased, the spatial profiles of the eigenmodes demonstrate clear changes, which can be described as a widening and merging of the neighboring maxima of the standing waves, resulting in a flattening of the standing-wave profile. In the limit of strong nonlinearity the distributions of the both shown modes trend toward very similar distributions, which are characterized by the reduced spatial periodicity in the longitudinal direction. The Fourier spectra for this strongly nonlinear case also demonstrate clear peaks at $k_z=\pi/w$ instead of $k_z=2\pi/w$ and $4\pi/w$ as found in the linear case. Such a modification can be understood as a nonlinear generation of additional long-wavelength spatial harmonics. Note here, that the studied system is characterized by the attractive nonlinearity [25], which, on the contrary, should lead to a preferred generation of short-wavelength spatial harmonics resulting in narrowing of the maxima of the standing waves and formation of two-dimensional soliton-like profiles from individual antinodes [26]. Therefore, one has to conclude, that the consideration of nonlinear spin-wave modes in in-plane magnetized confined magnetic objects demands taking into account other nonlinear phenomena in addition to the usually considered nonlinear modification of the spin-wave spectrum.

Recently similar reversed action of the spin-wave nonlinearity was observed in [27] for spin waves in the time domain. In this work it was shown that a time-domain standing wave formed by two spin waves can evolve into a sequence of dark solitons with increasing spin-wave amplitude. Note that, despite dark solitons usually appear in systems with the repulsive nonlinearity, in [27] they were observed for the spin-wave propagation geometry characterized by the attractive nonlinearity. This fact was explained by the action of the so-called nonlinear damping [28,29] which should be added to the Gilbert damping for proper description of the phenomenon. The nonlinear damping is caused by the parametric excitation of spin waves (see, e.g., [30]). Since the intensity of parametric processes strongly increases with increasing precession angle, this additional channel of the energy dissipation does not influence the magnetization dynamics for relatively small excitation powers, but can become significant if the power increases.

The very special feature of the nonlinear damping is that the parametric processes do not lead to a direct energy transfer from spin waves into the crystalline lattice. Instead, the energy is transferred into other spin waves characterized by very large wave vectors just due to the large phase volume of those waves [31]. Spin waves with large wave vectors are badly detectable with standard microwave techniques. However, excitation of such waves causes a reduction of the static magnetization of the medium. Therefore, the parametric excitation of spin waves with large wave vector can be detected through the frequency shift of spin-wave modes with small wave vectors, whose frequency depends on the static magnetization. Figure 3 demonstrates the dependences of the frequencies of the eigenmodes (2,1) and (4,1) as a function of the excitation power. For comparison, the average BLS intensity over the area of the square, which is a measure of the precession angle in the corresponding modes, is also presented in Fig. 3. As seen from the figure, the frequencies of the eigenmodes decrease linearly with increasing power indicating the linear reduction of the static magnetization $\Delta M_z$. Due to the conservation of the absolute value of the magnetization vector in the precessional dynamics, $\Delta M_z$ is proportional to the sum of the squared opening angles corresponding to different spin waves: $\Delta M_z \propto \sum_k |\varphi_k|^2$, where the summation takes into account all excited spin waves, including those not contributing to the BLS intensity. On the contrary, the average BLS intensity demonstrates a nonlinear behavior, indicating that with increasing precession angle the energy absorbed by the spin-wave eigenmode from the pumping is transferred into secondary spin waves more and more effectively. This fact can be considered as the experimental manifestation of the nonlinear damping effect.

**III. THEORY**

To model the magnetic eigenmodes of the magnetic square we use a dynamic equation derived in Appendix:

$$i\partial/\partial t\, F_{n.m}[\varphi(y,z)] + (\omega_{n,m} + i\eta - \omega)F_{n.m}[\varphi(y,z)] + (T + i\tau)F_{n.m}[|\varphi(y,z)|^2 \varphi(y,z)] = f_{n.m} \quad (1)$$

Here $y$ and $z$ are the in-plane coordinates aligned perpendicularly and parallel to the applied static magnetic field $H$, respectively. Function $\varphi(y,z)$ represents the spatial distribution of the dynamic magnetization in terms of the angle of magnetization precession. $F_{n.m}[\varphi(y,z)]$ and $F_{n.m}[|\varphi(y,z)|^2 \varphi(y,z)]$ are the amplitudes of the ($n,m$) spatial Fourier harmonics of the corresponding functions, i.e. $F_{n.m}[\varphi(y,z)]$ represents the amplitude of the linear eigenmode with indexes ($n,m$). $\omega_{n,m}$ is the frequency of the corresponding eigenmode in the linear regime. $\omega$ is the frequency of the driving microwave field produced by the wire antenna and $f_{n.m}$ are the Fourier amplitudes of this field. The nonlinear shift of the spin-wave spectrum is characterized by the nonlinear coefficient $T$, and the coefficients $\eta$ and $\tau$ describe the linear and the nonlinear damping, respectively [28].

Equation (1) is a confined-geometry analogue of the Complex Ginzburg-Landau Equation (CLGE) [27] written down in the Fourier space. In absence of linear and nonlinear damping, i.e. $\eta=\tau=0$, Eq. (1) is equivalent to the Fourier-space presentation of the two-dimensional Nonlinear Schroedinger Equation (NSE), which is widely used for description of propagation of nonlinear spin waves and spin-wave packets in an unbounded lossless magnetic medium [25]. Note here that, as seen from the analysis of the experimental data, the nonlinear damping is a key factor, which is in charge for the observed modifications of the eigenmode profiles. Therefore, taking it into account is absolutely necessary for the proper theoretical description of the phenomenon.

We solved the differential equation (1) on a mesh containing 64×64 cells by using the 4$^{th}$-order Runger-Kutta method combined with the Fast Fourier Transform algorithm. With the zero initial condition the method demonstrated convergence to stationary distributions $\varphi(y,z)$ for times smaller than $3/\eta$. For the coefficients, the values typical for YIG films were taken from the literature (see, e.g., [28]): $T$=10$^{10}$ rad/s, $\tau$=10$^{10}$ rad/s, and $2\eta/\gamma$=2$\Delta H$=0.5 Oe, where $\gamma = 2\pi \cdot 2.82 \cdot 10^6$ rad/(s·Oe) and $2\Delta H$ is the full ferromagnetic resonance linewidth. The frequencies of the linear eigenmodes $\omega_{n,m}$ were calculated based on the theory for spin waves in in-plane magnetized films developed in [32] and the quantized wave vectors $k_z$=$n\pi/w$, $k_y$=$m\pi/w$, where $w$=2 mm is the width of the magnetic element. As shown in [23] this method gives very good agreement with the experimentally measured frequencies.

In order to demonstrate the major role of the nonlinear damping in the observed phenomena at large excitation powers, we first made calculations of the nonlinear magnetization distributions based on Eq. (1) neglecting the nonlinear damping ($\tau$=0). The results of calculations are presented in Fig. 4 showing the obtained spatial distributions $|\varphi(y,z)|$ and their Fourier spectra for the strongly nonlinear case $\varphi$=2.8°. As seen from Fig. 4, in absence of the nonlinear damping the nonlinearity does not lead to a generation of long-wavelength harmonic with the wavevector $\pi/w$, as observed in the experiment. Instead, the energy is fed into a number of short-wavelength harmonics resulting in the appearance of the strongly localized maxima of the spin-wave intensity in the real space.

The results of calculations for the eigenmodes (2,1) and (4,1) taking into account the nonlinear damping are presented in Figs. 5 and 6, respectively. The panels (a)-(f) of the figures show the distributions $|\varphi(y,z)|$ and their two-dimensional Fourier spectra for different values of the mean angle $\varphi$ of the magnetic precession indicated in Figs. 5 and 6 close to the corresponding panels. As seen from Figs. 5 and 6, the results of calculations demonstrate the modification of the eigenmode profiles very similar to that observed in the experiments. As the precession angle increases from the value of 0.01° corresponding to the linear regime to the value of 2.8°, which is large enough for the response of the magnetic system to become essentially nonlinear (see, e.g., [33]), the Fourier spectra demonstrate gradual decrease of the amplitudes of the peaks corresponding to $k_z=2\pi/w$ and $4\pi/w$ and simultaneous appearance and the growth of the peaks for $k_z=\pi/w$. Finally, at the largest precession angle the Fourier components at $k_z=\pi/w$ dominate, which is in agreement with the experimental findings.

The observed modifications can be qualitatively understood based on an analysis of Eq. (1). In the limit of small amplitudes of dynamic magnetization $|\varphi(y,z)| \ll 1$ one can neglect the third term in the equation, which is in charge for the nonlinearity in the system. In this case the equation describes the process of linear excitation of the eigenmodes by a dynamic magnetic field at a fixed frequency $\omega$. The coefficient $(\omega_{n,m}+i\eta-\omega)$ in the second term of Eq. (1) determines the resonant response of the system if the frequency $\omega$ approaches the frequency of one of the Fourier harmonics $\omega_{n,m}$. As the system is excited at one of these resonant frequencies, the amplitude of the corresponding spatial Fourier harmonic $F_{n,m}[\varphi(y,z)]$ strongly dominates over others.

Therefore, in the linear case the spatial distributions of the dynamic magnetization of the eigenmodes represent clearly defined two-dimensional harmonic standing waves.

As the amplitude of the dynamic magnetization increases, the third term starts to play a significant role and a cross coupling between spatial harmonics appears. That means that, even if the system is excited at a frequency of one of the harmonics, the amplitudes of the other out-of-resonance harmonics become not vanishing and spatial distributions of the dynamic magnetization represent a combination of many standing waves with different spatial periods. This process can also be understood as a nonlinear energy channeling between different eigenmodes. Depending on the values of the nonlinear coefficient $T$ and the coefficient of the nonlinear damping $\tau$, the nonlinear coupling can channel the energy to out-of-resonance harmonics with either larger or smaller quantization indexes.

## IV. CONCLUSION

We have studied experimentally and theoretically spin-wave eigenmodes of saturated monodomain magnetic squares for different excitation powers. Our findings show that the nonlinearity in the system leads to a significant modification of the spatial profiles of dynamic magnetization, which can be understood as the nonlinear generation of long-wavelength spatial harmonics caused by the nonlinear coupling between different eigenmodes and the energy channeling between them. The experimental observations are well reproduced in the theoretical calculations taking into account the nonlinear spin-wave damping. The observed phenomenon of the nonlinear mode coupling can be used, for example, for nonlinear amplification and/or reduction of magnetic losses of the eigenmodes.

**ACKNOWLEDGMENTS**

This work was supported in part by the Deutsche Forschungsgemeinschaft and the Australian Research Council (ARC).

**APPENDIX**

Here we show derivation of Eq. (1) describing nonlinear spin dynamics in rectangular magnetic elements. We start with an integral equation for plain spin waves propagating in a continuos magnetic film derived in [34]. It has the form as follows:

$$\chi(z,\omega)^{-1} m_x(z) - \int_{-\infty}^{+\infty} 4\pi G_{xx}(z,z') m_x(z') dz' = h_x(z). \tag{A1}$$

Here $z$ is the propagation coordinate, coinciding with the direction of the static magnetic field $H$, and the coordinate $x$ is oriented normally to the film plane. $m_x$ is the out-of-plane component of the dynamic magnetization, $h_x$ is the dynamic magnetic field of the exciting antenna having frequency $\omega$, $\chi$ is the diagonal element of the microwave permeability tensor of a ferromagnet [35]

$$\chi = \frac{1}{4\pi} \frac{\omega_H \omega_M}{\omega_H^2 - \omega^2}, \tag{A2}$$

$G_{xx}$ is the diagonal component of the quasi-one-dimensional magnetostatic Green's function [5]

$$4\pi G_{xx}(z,z') = \frac{2}{L} \ln \frac{(z-z')^2}{L^2 + (z-z')^2}, \tag{A3}$$

$\omega_H = \gamma H$, $\omega_M = \gamma 4\pi M_z$, $\gamma = 2\pi \cdot 2.82 \cdot 10^6$ rad/(s·Oe), and $L$ is the film thickness.

Note, that the quantity $\omega_M$ is $z$–dependent, because the longitudinal component $M_z$ of the full magnetization vector is $z$-dependent due to the nonlinearity of spin waves:

$$M_z(t,z) = \left\{ M_S^2 - \left[\mathrm{Re}(m_x(t,z))\right]^2 - \left[\mathrm{Re}(m_y(t,z))\right]^2 \right\}^{1/2}, \qquad (A4)$$

where $M_s$ is the saturation magnetization and $m_x$ and $m_y$ are the magnitudes of transverse components of the dynamic magnetization. These components are connected through the microwave permeability tensor as follows:

$$m_y = i\varepsilon m_x, \qquad (A5)$$

where $\varepsilon = \omega/\omega_H$ is the ellipticity factor. Substituting (A5) and the time dependence of dynamic magnetization in the form $m_x \exp(i\omega t)$ into (A4) we arrive at the expression for the value of $M_z$ averaged over the precession period $2\pi/\omega$:

$$M_z \approx M_S - (1+\varepsilon^2)/4 |m_x|^2. \qquad (A6)$$

Here we used the fact that

$$|m_x|^2 \ll M_z^2. \qquad (A7)$$

Inserting (A6) into (A2) and using once again the condition (A7) we get:

$$\chi(z,\omega)^{-1} = v^{(0)} + v^{(2)} |m_x(z)|^2, \qquad (A8)$$

where $v^{(0)}$ is the linear inverse microwave permeability which coincides with $\chi^{-1}$ from (A2) with $\omega_M = \gamma 4\pi M_S$, and $v^{(2)}$ is the inverse nonlinear microwave permeability:

$$v^{(2)} = v^{(0)}(1+\varepsilon^2)/(16\pi M_S). \qquad (A9)$$

With (A8) Eq. (A1) reduces to:

$$v^{(0)} m_x(z) - \int_{-\infty}^{+\infty} 4\pi G_{xx}(z,z') m_x(z') dz' + v^{(2)} |m_x(z)|^2 m_x(z) = h_x(z). \qquad (A10)$$

Now we introduce confinement in the $z$-direction. We assume that the magnetic element has the length $w$. It was previously shown that the precession amplitude at the lateral edges of magnetic elements is significantly smaller than in the middle of the sample [5], i.e. the so-called effective pinning of dynamic magnetization at the edges takes place. It was also shown that this behavior continues to the nonlinear regime of spin-wave waveform propagation [36, 37].

The function $\sin(kz) = [\exp(ikz) - \exp(-ikz)]/(2i)$ represents the eigenfunction of the integral operator $\int_{-\infty}^{+\infty} G_{xx}(z,z')dz'$ [34]. The corresponding eigenvalue is $W(k) = \frac{1}{|k|L}\left[\exp(-|k|L) - 1\right]$. In [5,15] it was shown that $W(k)$ values for discrete wave vectors $k_n$ satisfying the effective boundary conditions for the harmonic function $A\sin(k_n z) + B\cos(k_n z)$ represent good approximation to the eigenvalues of the integral operator with a finite range of integration: $\int_0^w G_{xx}(z,z')dz'$ and the functions $A\sin(k_n z) + B\cos(k_n z)$ represent good approximations for the eigenfunctions of this operator. Furthermore, it was previously shown [36] that for qualitative description of nonlinear effects it is sufficient to consider the dynamic magnetization at the sample edges as totally pinned: $m_x(z=0) = 0$, $m_x(z=w) = 0$, then $B = 0$ and $k_n = n\pi/w$. Thus

$$\int_0^w G_{xx}(z,z')\sin(k_n z')dz' \approx W(k)\sin(k_n z). \qquad (A11)$$

Since the set of functions $\sin(k_n z)$ is a full orthogonal set of functions, the profile of the dynamic magnetization in (A10) can be expanded in sine Fourier series:

$$m_x(z) = \sum_{n=1}^{\infty} m_n \sin(k_n z). \tag{A12}$$

Then substituting (A11) into (A10) and using the orthogonality condition for the sine functions we get:

$$\left[ v^{(0)} - 4\pi W(k) \right] m_n + \frac{v^{(2)}}{w} \int_0^w |m_x(z)|^2 m_x(z) \sin(k_n z) dz = h_n. \tag{A13}$$

Here $m_n = \frac{1}{w} \int_0^w m_x(z) \sin(k_n z) dz$ and $h_n = \frac{1}{w} \int_0^w h_x(z) \sin(k_n z) dz$. Using (A2) the quantity $v^{(0)}/(4\pi) - W(k)$ can be rewritten as $(\omega_n^2 - \omega^2)/(\omega_H \omega_M)$, where $\omega_n^2 = \omega_H(\omega_H + \omega_M) - \omega_H \omega_M [1 + W(k)]$. Since for $h_x = 0$ and $|m_x(z)|^2 \to 0$ (A13) reduces to $\left[ v^{(0)}/(4\pi) - W(k) \right] m_n = 0$, the frequencies $\omega_n$ represent eigenfrequencies of the linear eigenmodes of the structure.

The effect we are considering actually represents an enrichment of the spectrum of resonance oscillations due to a weak nonlinearity. Generation of new harmonics is efficient if their linear frequency shift from the main harmonic $\omega_{n0}$ is of the order of the nonlinear frequency shift. Therefore all eigenfrequencies $\omega_n$ are close to $\omega_{n0}$ which in its turn is close to the frequency of the driving field $\omega$ (and equals to $\omega$ in the linear regime). Then for all $\omega_{n0}$ of the nonlinear standing-wave package it stands $\omega_n^2 - \omega^2 = (\omega_n - \omega)(\omega + \omega_n) \approx 2\omega(\omega_n - \omega)$ and we arrive to a formula as follows:

$$(\omega_n - \omega)\varphi_n + T' \int_0^w |\varphi(z)|^2 \varphi(z) \sin(k_n z) dz = f_n, \tag{A14}$$

where

$$T' = \frac{\omega_H \omega_M M_S}{16 w \pi \omega} v^{(0)}, \tag{A15}$$

$f_n = \dfrac{\omega_H \omega_M}{\sqrt{2} 8\pi \omega_n M_S} h_n$, and $\varphi = \sqrt{(1+\varepsilon^2)/2}\, m_x / M_S$ is the mean precession angle.

Finally we use the usual substitution $\omega \to \omega - i\partial/\partial t$ to describe time evolution of the slow envelope and extend the formula onto the two-dimensional case of a square magnetic element with the area $w \times w$:

$$i\partial/\partial t\, F_{n.m}[\varphi(y,z)] + (\omega_{n,m} + i\eta - \omega) F_{n.m}[\varphi(y,z)] + (T + i\tau) F_{n.m}\left[|\varphi(y,z)|^2 \varphi(y,z)\right] = f_{n.m}.$$

(A16)

In this formula $F_{n,m}[u(y,z)]$ denotes the two-dimensional sine Fourier transform of a function $u(y,z)$, and $T = T'w$. The additional terms in this expression account for the intensity-dependent spin wave damping $\eta + \tau |\varphi(y,z)|^2$ with $\eta, \tau > 0$. They are introduced phenomenologically, as suggested in [28].

# REFERENCES

[1] C. Mathieu, J. Jorzick, A. Frank, S.O. Demokritov, A.N. Slavin, and B. Hillebrands , Phys. Rev. Lett. **81,** 3968 (1998).

[2] J. Jorzick, S.O. Demokritov, C. Mathieu, B. Hillebrands, B. Bartenlian, C. Chappert, F. Rousseaux, A.N. Slavin, Phys. Rev. B **60,** 15194 (1999).

[3] J. Jorzick, S.O. Demokritov, B. Hillebrands, B. Bartenlian, C. Chappert, D. Decanini, F. Rousseaux, and E. Cambril, Appl. Phys. Lett. **75**, 3859 (1999).

[4] J. Jorzick, S.O. Demokritov, B. Hillebrands, M. Bailleul, C. Fermon, K.Y. Guslienko, A.N. Slavin, D.V. Berkov, and N.L. Gorn, Phys. Rev. Lett. **88**, 047204 (2002).

[5] K.Yu. Guslienko, S.O. Demokritov, B. Hillebrands, and A.N. Slavin, Phys. Rev. B, **66**, 132402 (2002).

[6] J.P. Park, P. Eames, D.M. Engebretson, J. Berezovsky, and P.A. Crowell, Phys. Rev. B **67,** 020403(R) (2003).

[7] T.M. Crawford, M. Covington, and G.J. Parker, Phys. Rev. B **67**, 024411 (2003).

[8] G. Gubbiotti, G. Carlotti, T. Okuno, T. Shinjo, F. Nizzoli, and R. Zivieri, Phys. Rev. B **68**, 184409 (2003).

[9] M. Grimsditch, G.K. Leaf, H.G. Kaper, D.A. Karpeev, and R.E. Camley, Phys. Rev. B **69**, 174428 (2004).

[10] M. Buess, R. Höllinger, T. Haug, K. Perzlmaier, U. Krey, D. Pescia, M.R. Scheinfein, D. Weiss, and C.H. Back , Phys. Rev. Lett. **93**, 077207 (2004).

[11] B.A. Ivanov and C.E. Zaspel, Phys. Rev. Lett. **94**, 027205 (2005).

[12] K. Perzlmaier, M. Buess, C.H. Back, V.E. Demidov, B. Hillebrands, S.O. Demokritov, Phys. Rev. Lett. **94,** 057202 (2005).

[13] J. Raabe, C. Quitmann, C.H. Back, F. Nolting, S. Johnson, and C. Buehler, Phys. Rev. Lett. **94**, 217204 (2005).

[14] J.P. Park and P.A. Crowell, Phys. Rev. Lett. **95**, 167201 (2005).

[15] C. Bayer, J. Jorzick, B. Hillebrands, S.O. Demokritov, R. Kouba, R. Bozinoski, A.N. Slavin, K.Y. Guslienko, D.V. Berkov, N.L. Gorn, and M.P. Kostylev, Phys. Rev. B **72**, 064427 (2005).

[16] J. Podbielski, F. Giesen, and D. Grundler, Phys. Rev. Lett. **96**, 167207 (2006).

[17] M. Bailleul, R. Höllinger, and C. Fermon, Phys. Rev. B **73**, 104424 (2006).

[18] I. Neudecker, M. Kläui, K. Perzlmaier, D. Backes, L.J. Heyderman, C.A.F. Vaz, J.A.C. Bland, U. Rüdiger, and C.H. Back, Phys. Rev. Lett. **96**, 057207 (2006).

[19] M.P. Kostylev, G. Gubbiotti, J.-G. Hu, G. Carlotti, T. Ono, and R.L. Stamps, Phys. Rev. B, **76**, 054422 (2007).

[20] G. Gubbiotti, S. Tacchi, G. Carlotti, N. Singh, S. Goolaup, A. O. Adeyeye, and M. Kostylev, Appl. Phys. Lett. **90**, 092503 (2007).

[21] J.C. Sankey, I.N. Krivorotov, S.I. Kiselev, P.M. Braganca, N.C. Emley, R.A. Buhrman, and D.C. Ralph, Phys. Rev. B **72**, 224427 (2005).

[22] J.C. Sankey, P.M. Braganca, A.G.F. Garcia, I.N. Krivorotov, R.A. Buhrman, and D.C. Ralph, Phys. Rev. Lett. **96**, 227601 (2006).

[23] V.E. Demidov, U.-H. Hansen, and S.O. Demokritov, Phys. Rev. Lett. **98**, 157203 (2007).

[24] S.O. Demokritov, B. Hillebrands, and A.N. Slavin, Phys. Rep. **348**, 441 (2001).

[25] A. Zvezdin and A. Popkov, JETP **57**, 350 (1983).

[26]   A. Slavin and V. Tiberkevich, Phys. Rev. Lett. **95**, 237201 (2005).

[27]   M.M. Scott, M.P. Kostylev, B.A. Kalinikos, C.E. Patton, Phys. Rev. B **71**, 174440 (2005).

[28]   M.M. Scott, C.E. Patton, M.P. Kostylev, and B.A. Kalinikos, J. Appl. Phys, **95**, 6294 (2004).

[29]   V. Tiberkevich and A. Slavin, Phys. Rev. B **75**, 014440 (2007).

[30]   K. Livesey, R.L. Stamps, and M. Kostylev, Phys. Rev. B **75**, 174427 (2007).

[31]   V.E. Demidov, O. Dzyapko, S.O. Demokritov, G. A. Melkov, and A. N. Slavin, Phys. Rev. Lett. **99**, 037205 (2007).

[32]   R.W. Damon and J.R. Eshbach, J. Phys. Chem. Solids **19**, 308 (1961).

[33]   H.Y. Zhang, P. Kabos, H. Xia, R.A. Staudinger, P.A. Kolodin, and C.E. Patton, J. Appl. Phys. **84**, 3776 (1998).

[34]   M.P. Kostylev, A.A. Serga, T. Schneider, B. Leven, B. Hillebrands, and R.L. Stamps, Phys. Rev. B (to be published).

[35]   A.G. Gurevich and G.A. Melkov, "*Magnetization oscillations and waves*" (CRC Press, New York, 1996).

[36]   A.A. Serga, M. Kostylev, and B. Hillebrands, arXiv:0704.0024 (http://arxiv.org/abs/0704.0024 ); M. Kostylev, A.A. Serga, and B. Hillebrands, "Two dimensional microwave nonlinear spin-wave pulses in in-plane confined magnetic films", Technical Digests of International Magnetic Conference (Intermag'2006), May 8-12, 2006, San Diego, California, FV03 (2006).

[37]   V.E. Demidov, U.-F. Hansen, O. Dzyapko, N. Koulev, S.O. Demokritov, and A.N. Slavin, Phys. Rev. B **74**, 092407 (2006).

**FIGURE CAPTIONS**

FIG. 1. (Color online) Frequency dependence of the microwave absorption by the square magnetic element for $H$=800 Oe and $P$=1 mW. The resonant peaks appear at the frequencies of the eigenmodes and are labeled with two indexes ($n,m$) corresponding to the number of antinodes in the standing waves in the directions parallel and perpendicular to the static magnetic field, respectively

FIG. 2. (Color online) Measured distributions of the dynamic magnetization and their two-dimensional spatial Fourier spectra for the eigenmodes (2,1) (a) and (4,1) (b) in the linear and strongly nonlinear regimes corresponding to the excitation power equal to 1 and 200 mW, respectively.

FIG. 3. (Color online) Experimental dependences of the frequencies of the eigenmodes (2,1) and (4,1) and the average BLS intensity obtained from the corresponding two-dimensional distributions on the excitation power.

FIG. 4. (Color online) Distributions of the dynamic magnetization and their two-dimensional Fourier spectra calculated neglecting the nonlinear damping ($\tau$=0) for the eigenmodes (2,1) (a) and (4,1) (b). The distributions correspond to the strongly nonlinear case $\varphi$=2.8°.

FIG. 5. (Color online) Distributions of the dynamic magnetization and their two-dimensional Fourier spectra for the eigenmode (2,1) calculated taking into account the nonlinear damping. The distributions are characterized by the mean precession angle $\varphi$, indicated close to the corresponding panels.

FIG. 6. (Color online) Distributions of the dynamic magnetization and their two-dimensional Fourier spectra for the eigenmode (4,1) calculated taking into account the nonlinear damping.

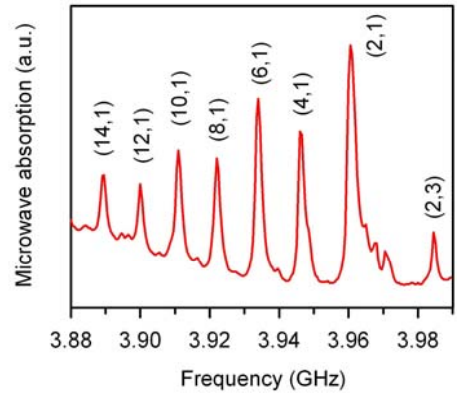

FIG. 1

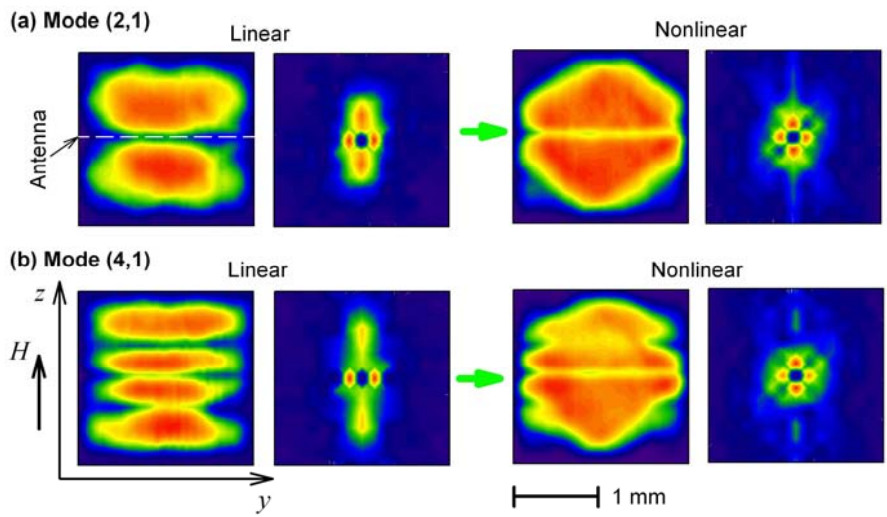

FIG. 2

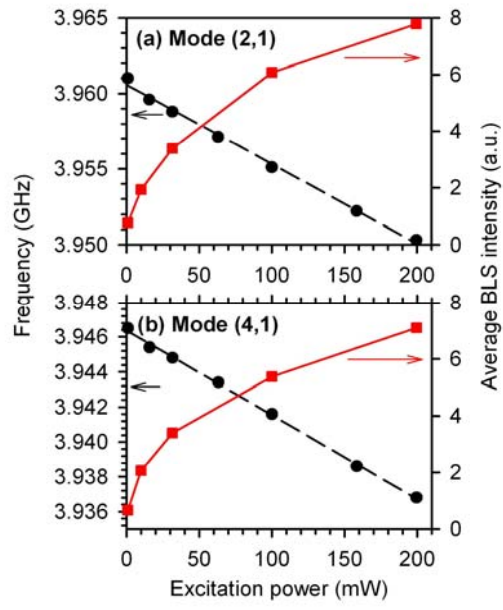

FIG. 3

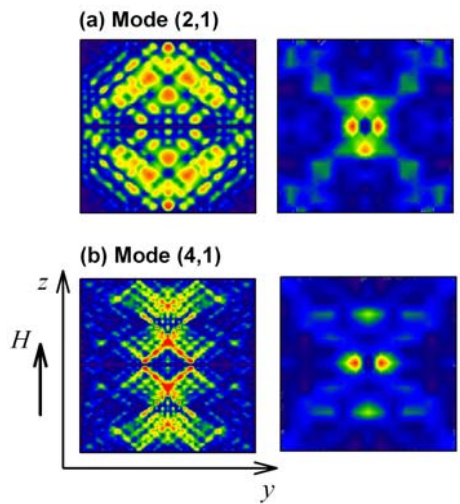

FIG. 4

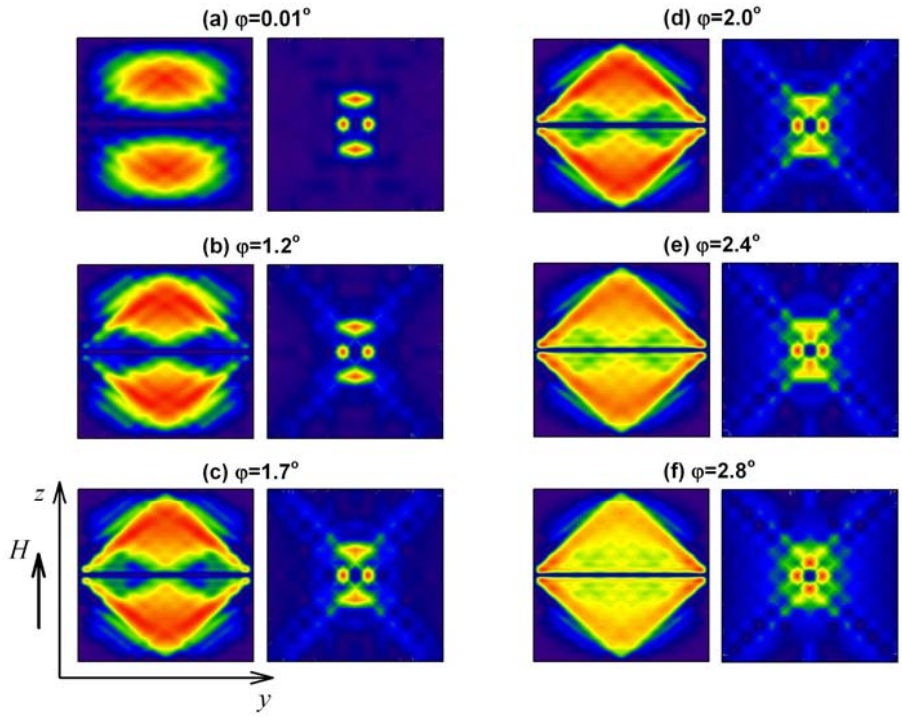

FIG. 5

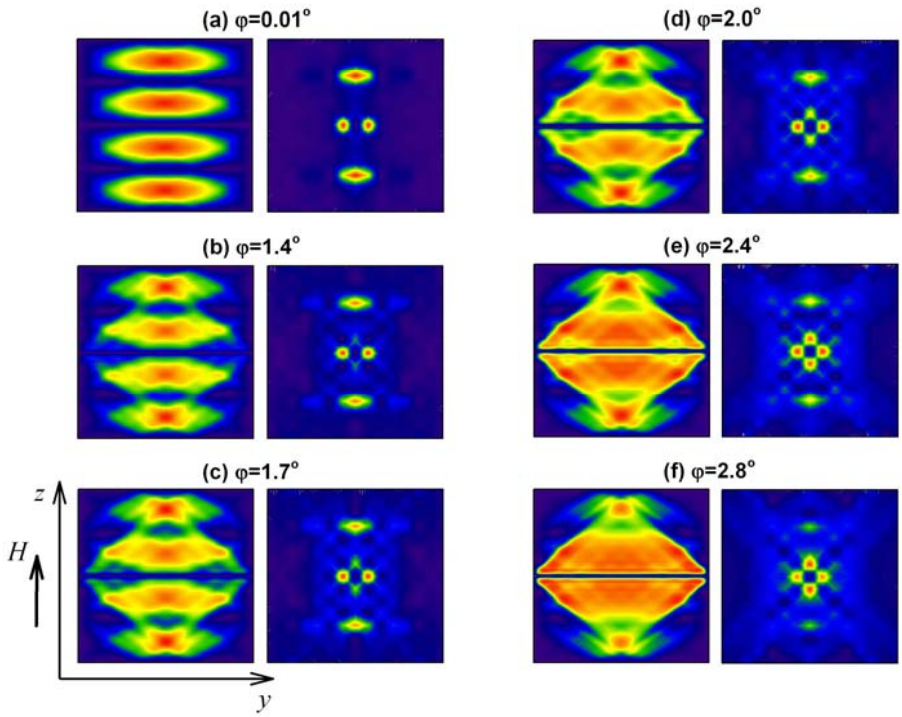

FIG. 6